\documentclass[]{aastex701}

\usepackage{subfiles}
\usepackage{multirow}
\usepackage{longtable}
\usepackage{rotating}
\usepackage{adjustbox}
\usepackage{xspace}
\usepackage{amsfonts}
\usepackage{subcaption}
\usepackage{caption}

\newcommand{\HeI}{He~{\sc i}}
\newcommand{\NaI}{Na~{\sc i}}
\newcommand{\FeII}{Fe~{\sc ii}}
\newcommand{\SiII}{Si~{\sc ii}}
\newcommand{\SII}{S~{\sc ii}}
\newcommand{\CaII}{Ca~{\sc ii}}
\newcommand{\STag}{\texttt{STag}\xspace}
\newcommand{\STagII}{\texttt{STag~II}\xspace}

\renewcommand{\thefootnote}{\alph{footnote}}

\graphicspath{{./}{figures/}}

\begin{document}

\title{STag II: Classification of Serendipitous Supernovae Observed by Galaxy Redshift Surveys}


\author{W.~Davison}
\affiliation{Korea Astronomy and Space Science Institute, 776, Daedeokdae-ro, Yuseong-gu, Daejeon 34055, Republic of Korea}
\affiliation{University of Science and Technology, 217, Gajeong-ro, Yuseong-gu, Daejeon 34113, Republic of Korea}
\email[show]{william.davison117@gmail.com}

\author[orcid=0000-0002-7464-2351]{D.~Parkinson}
\affiliation{Korea Astronomy and Space Science Institute, 776, Daedeokdae-ro, Yuseong-gu, Daejeon 34055, Republic of Korea}
\affiliation{University of Science and Technology, 217, Gajeong-ro, Yuseong-gu, Daejeon 34113, Republic of Korea}
\email[show]{davidparkinson@kasi.re.kr}

\author[orcid=0000-0001-5537-4710]{S.~BenZvi}
\affiliation{Department of Physics \& Astronomy, University of Rochester, 206 Bausch and Lomb Hall, P.O. Box 270171, Rochester, NY 14627-0171, USA}
\email{sybenzvi@pas.rochester.edu}

\author{A.~Palmese}
\affiliation{Department of Physics, Carnegie Mellon University, 5000 Forbes Avenue, Pittsburgh, PA 15213, USA}
\email{antonella.palmese@gmail.com}
\author{J.~Aguilar}
\affiliation{Lawrence Berkeley National Laboratory, 1 Cyclotron Road, Berkeley, CA 94720, USA}
\email{jaguilar@lbl.gov}
\author[orcid=0000-0001-6098-7247]{S.~Ahlen}
\affiliation{Physics Dept., Boston University, 590 Commonwealth Avenue, Boston, MA 02215, USA}
\email{ahlen@bu.edu}
\author{D.~Brooks}
\affiliation{Department of Physics \& Astronomy, University College London, Gower Street, London, WC1E 6BT, UK}
\email{david.brooks@ucl.ac.uk}
\author{T.~Claybaugh}
\affiliation{Lawrence Berkeley National Laboratory, 1 Cyclotron Road, Berkeley, CA 94720, USA}
\email{tmclaybaugh@lbl.gov}
\author[orcid=0000-0002-1769-1640]{A.~de la Macorra}
\affiliation{Instituto de F\'{\i}sica, Universidad Nacional Aut\'{o}noma de M\'{e}xico,  Cd. de M\'{e}xico  C.P. 04510,  M\'{e}xico}
\email{macorra@fisica.unam.mx}
\author[orcid=0000-0002-4928-4003]{Arjun~Dey}
\affiliation{NSF NOIRLab, 950 N. Cherry Ave., Tucson, AZ 85719, USA}
\email{arjun.dey@noirlab.edu}
\author{P.~Doel}
\affiliation{Department of Physics \& Astronomy, University College London, Gower Street, London, WC1E 6BT, UK}
\email{apd@star.ucl.ac.uk}
\author{E.~Gaztañaga}
\affiliation{Institut d'Estudis Espacials de Catalunya (IEEC), 08034 Barcelona, Spain}
\affiliation{Institute of Cosmology and Gravitation, University of Portsmouth, Dennis Sciama Building, Portsmouth, PO1 3FX, UK}
\affiliation{Institute of Space Sciences, ICE-CSIC, Campus UAB, Carrer de Can Magrans s/n, 08913 Bellaterra, Barcelona, Spain}
\email{}
\author[orcid=0000-0003-3142-233X]{S.~Gontcho A Gontcho}
\affiliation{Lawrence Berkeley National Laboratory, 1 Cyclotron Road, Berkeley, CA 94720, USA}
\email{}
\author[orcid=0000-0002-1081-9410]{C.~Howlett}
\affiliation{School of Mathematics and Physics, University of Queensland, 4072, Australia}
\email{}
\author{S.~Juneau}
\affiliation{NSF NOIRLab, 950 N. Cherry Ave., Tucson, AZ 85719, USA}
\email{}
\author[orcid=0000-0003-3510-7134]{T.~Kisner}
\affiliation{Lawrence Berkeley National Laboratory, 1 Cyclotron Road, Berkeley, CA 94720, USA}
\email{}
\author[orcid=0000-0001-6356-7424]{A.~Kremin}
\affiliation{Lawrence Berkeley National Laboratory, 1 Cyclotron Road, Berkeley, CA 94720, USA}
\email{}
\author{A.~Lambert}
\affiliation{Lawrence Berkeley National Laboratory, 1 Cyclotron Road, Berkeley, CA 94720, USA}
\email{}
\author[orcid=0000-0003-1838-8528]{M.~Landriau}
\affiliation{Lawrence Berkeley National Laboratory, 1 Cyclotron Road, Berkeley, CA 94720, USA}
\email{}
\author[orcid=0000-0001-7178-8868]{L.~Le~Guillou}
\affiliation{Sorbonne Universit\'{e}, CNRS/IN2P3, Laboratoire de Physique Nucl\'{e}aire et de Hautes Energies (LPNHE), FR-75005 Paris, France}
\email{}
\author[orcid=0000-0002-1125-7384]{A.~Meisner}
\affiliation{NSF NOIRLab, 950 N. Cherry Ave., Tucson, AZ 85719, USA}
\email{}
\author{R.~Miquel}
\affiliation{Instituci\'{o} Catalana de Recerca i Estudis Avan\c{c}ats, Passeig de Llu\'{\i}s Companys, 23, 08010 Barcelona, Spain}
\affiliation{Institut de F\'{i}sica d’Altes Energies (IFAE), The Barcelona Institute of Science and Technology, Campus UAB, 08193 Bellaterra Barcelona, Spain}
\email{}
\author[orcid=0000-0002-2733-4559]{J.~Moustakas}
\affiliation{Department of Physics and Astronomy, Siena College, 515 Loudon Road, Loudonville, NY 12211, USA}
\email{}
\author{A.~D.~Myers}
\affiliation{Department of Physics \& Astronomy, University  of Wyoming, 1000 E. University, Dept.~3905, Laramie, WY 82071, USA}
\email{}
\author{C.~Poppett}
\affiliation{Lawrence Berkeley National Laboratory, 1 Cyclotron Road, Berkeley, CA 94720, USA}
\affiliation{Space Sciences Laboratory, University of California, Berkeley, 7 Gauss Way, Berkeley, CA  94720, USA}
\affiliation{University of California, Berkeley, 110 Sproul Hall \#5800 Berkeley, CA 94720, USA}
\email{}
\author[orcid=0000-0001-7145-8674]{F.~Prada}
\affiliation{Instituto de Astrof\'{i}sica de Andaluc\'{i}a (CSIC), Glorieta de la Astronom\'{i}a, s/n, E-18008 Granada, Spain}
\email{}
\author[orcid=0000-0001-5589-7116]{M.~Rezaie}
\affiliation{Department of Physics, Kansas State University, 116 Cardwell Hall, Manhattan, KS 66506, USA}
\email{}
\author{G.~Rossi}
\affiliation{Department of Physics and Astronomy, Sejong University, Seoul, 143-747, Korea}
\email{}
\author[orcid=0000-0002-9646-8198]{E.~Sanchez}
\affiliation{CIEMAT, Avenida Complutense 40, E-28040 Madrid, Spain}
\email{}
\author[orcid=0000-0002-3569-7421]{E.~F.~Schlafly}
\affiliation{Space Telescope Science Institute, 3700 San Martin Drive, Baltimore, MD 21218, USA}
\email{}
\author{M.~Schubnell}
\affiliation{Department of Physics, University of Michigan, Ann Arbor, MI 48109, USA}
\affiliation{University of Michigan, Ann Arbor, MI 48109, USA}
\email{}
\author{D.~Sprayberry}
\affiliation{NSF NOIRLab, 950 N. Cherry Ave., Tucson, AZ 85719, USA}
\email{}
\author[orcid=0000-0003-1704-0781]{G.~Tarl\'{e}}
\affiliation{University of Michigan, Ann Arbor, MI 48109, USA}
\email{}
\author{B.~A.~Weaver}
\affiliation{Lawrence Berkeley National Laboratory, 1 Cyclotron Road, Berkeley, CA 94720, USA}
\email{}
\author[orcid=0000-0002-6684-3997]{H.~Zou}
\affiliation{National Astronomical Observatories, Chinese Academy of Sciences, A20 Datun Rd., Chaoyang District, Beijing, 100012, P.R. China}
\email{}


\begin{abstract}

With the number of supernovae observed expected to drastically increase thanks to large-scale surveys like the Dark Energy Spectroscopic Instrument (DESI), it is necessary that the tools we use to classify these objects keep up with this increase. We previously created \textit{Supernova Tagging and Classification} (\STag) to address this problem by employing machine learning techniques alongside logistic regression in order to assign `tags' to spectra based on spectral features. \STagII is a continuation of this work, which now makes use of model supernova spectra combined with real DESI spectra in order to train \STag to better deal with realistic data. We also make use of the \textit{r}lap score as a trustworthiness cut, making for a more robust and accurate supernova classifier than before.
\end{abstract}


\section{Introduction}\label{sec:intro}
The Dark Energy Spectroscopic Instrument (DESI) is a spectroscopic survey that will observe some 40 million galaxies \citep{2013arXiv1308.0847L,DESI2016a.Science,2016arXiv161100037D,DESI2022.KP1.Instr,2023AJ....165....9S,2023AJ....165..144G,2023arXiv230606310M,2023AJ....166..259S}. Whilst not a supernova (SN) survey itself, DESI will undoubtedly obtain the spectra for a significant quantity of SN, either through intentional follow-up of photometric surveys \citep{MostHosts}, or by serendipitous discovery, similar to the recent case with the Hobby–Eberly Telescope Dark Energy eXperiment (HETDEX; \citealt{vinko2023}). SN, specifically Type Ia, have been a crucial part of cosmology for a long time, for example as evidence for the acceleration of the expansion of the Universe \citep{Riess1998,Perlmutter1999} and for estimating the Hubble constant (H$_0$; \citealt{Riess2022}). More recently, attempts to use Type II SNe as standardisable candles have become increasingly popular \citep{Jaeger2022}. Current and future surveys (such as DESI) will produce many more spectra than can be analysed by typical manual methods \citep{Howell2005,Blondin2007} and so using machine learning to spectroscopically classify transients has become an area of growing interest \citep{Hala2014,Sasdelli2016,Muthukrishna2019,Davison2022}. 

In order to classify something spectroscopically we need to do feature extraction, which comes in a variety of flavours. One may look at the spectrum as a whole and use template spectra to do cross-correlation, as with \textit{Deep Automated SN and Host Classifier} (\texttt{DASH}; \citealt{Muthukrishna2019}). In our previous paper \citep{Davison2022} we introduced a new method which uses the concept of feature tags to classify SN spectra. This method, known as \textit{Supernova Tagging and Classification} (\STag), used logistic regression \citep{logistic_regression} to assign tag probabilities to distinct spectroscopic features; namely spectral lines. This is a process called multi-label classification, whereby each tag probability is calculated independently \citep{Read2011}. Each spectrum would have the same set of tags, but with different probabilities for each spectral line, and these tags are then passed to a simple feedforward neural network consisting of an input layer, one hidden layer, and an output layer which uses softmax regression to determine the most appropriate class based on the tag probabilities; all layers of \STag are fully connected. The artificial neural network (ANN; \citealt{Rumelhart1986}) is trained such that it learns to associate high (or low) tag probabilities of particular spectral lines with a particular SN class; for example we would expect a Type II SN to have a high H$\alpha$ tag probability, but other tag probabilities (high or low) may also have an unforeseen impact.

\STag was trained using template spectra, which are idealised representations of actual SN spectra. In reality, SN spectra often have contamination from their host galaxy in the form of spectral features, as well as continuum contamination since the continuum-removal process is done only by approximation (specifically, a cubic spline is used to model the continuum which the full spectra is then divided by). What this means in practice is that spectral features that are associated with the galaxy light also appear in SN spectra, and may fall at the same wavelength ranges as spectral features of the SN. As such it is imperative to develop a technique that can distinguish between the two, and so extract another feature from the data for use in classification. We make use of the equivalent width of a spectral line, whereby the strength and/or sign of the equivalent width (negative for absorption, positive for emission) can be used to identify whether a line is the desired feature or a contaminating one. We also note that different elemental spectral lines can occur at the same position in a spectra as other spectral lines, and that the equivalent width may be a way of distinguishing the two; such as with \HeI\ $\lambda$5876 and \NaI\ D $\lambda$5895 potentially having overlap at the same wavelength range, posing an issue for Type Ib vs. Type Ic supernovae (SNe) classifications \citep{Branch2002,Elmhamdi2006,Liu2016}. Furthermore, it has been shown that the strength of a specific spectral line can be used to distinguish between different classes, such as the case of H$\alpha$ for Type II and Type Ib SNe or O I $\lambda$7774 for Type Ib and Type Ic SNe \citep{Matheson2001,Liu2016}.

To that end, we present an updated version of \STag which utilises not only updated versions of previously made tags, but also the equivalent widths of each spectral line and a non-linear combination of both the tag probability and associated equivalent width. This version of \STag has demonstrably improved performance over the original version and has been used to classify new spectra. In follow-up work, this new version of \STag will be used to classify DESI spectra of transients identified by other machine learning methods. This paper is organised as follows: Section \ref{sec:data} details the data used for the training of \STag as well as the new spectra it has been used to classify. Section \ref{sec:meth} describes the general process by which \STag works, as well as how the equivalent widths are implemented. In Section \ref{sec:res} we showcase the classification of DESI SN and finally in Section \ref{sec:dis} we talk about the implications of this new version of \STag. 

\renewcommand{\thefootnote}{\arabic{footnote}}

\section{Data}\label{sec:data}
Whilst the first version of \STag made use of the full set of template spectra used in \texttt{DASH} \citep{Muthukrishna2019} and real data from the Australian Dark Energy Survey (OzDES; \citealt{Lidman2020}), we are now interested in adapting \STag for use with DESI. For this reason we instead use template spectra that also make use of real DESI data \citep{2023arXiv230606308D,2024AJ....167...62D}.

\subsection{Simulated Supernova DESI Spectra}\label{sec:sim}
Since DESI is not a survey designed to observe SN, the vast majority of DESI spectra will not be of a SN. As such, we use DESI spectra of real galaxies that have been combined with template SN spectra. These templates consist of both core-collapse SNe (Type Ib, Ic, and II) \citep{Vincenzi2019} and Type Ia SNe \citep{Hsiao2007}, allowing us to cover the full range of SN types we wish to be able to classify. We note here that this is a different set of supernova templates to those used in the first version of \STag \citep{Davison2022}. These models are first loaded into a script, with certain core-collapse models blacklisted due to being non-standard (namely SN 2013by, SN 2013fs, SN 2009bw, SN 2012aw, SN 2009kr, ASASSN14j, SN 2013am, SN 2008ax, SN 2008fq, SN 2009ip, iPTF13bvn, SN 2008D, SN 1994I, SN 2007gr, SN 2009b, and SN 2007ru). These spectra were not included for a variety of reasons, such as due to having narrower/broader lines \citep{Hunter2009,Sahu2009,Taddia2013,Black2017}, representing a type of sub-luminous Type II SNe \citep{Tomasella2018}, otherwise unique looking spectra \citep{Filippenko1995,Mazzali2008,Fraser2010,Inserra2012,Bose2013,Fremling2014,Yaron2017}, being a Type IIb SNe \citep{Pastorello2008}, or, in the case of SN 2009ip, ongoing discussion as to whether it is a true SN or just eruptions from a massive star \citep{Pessi2023}.

\begin{figure}[htbp]
\centering
\includegraphics[]{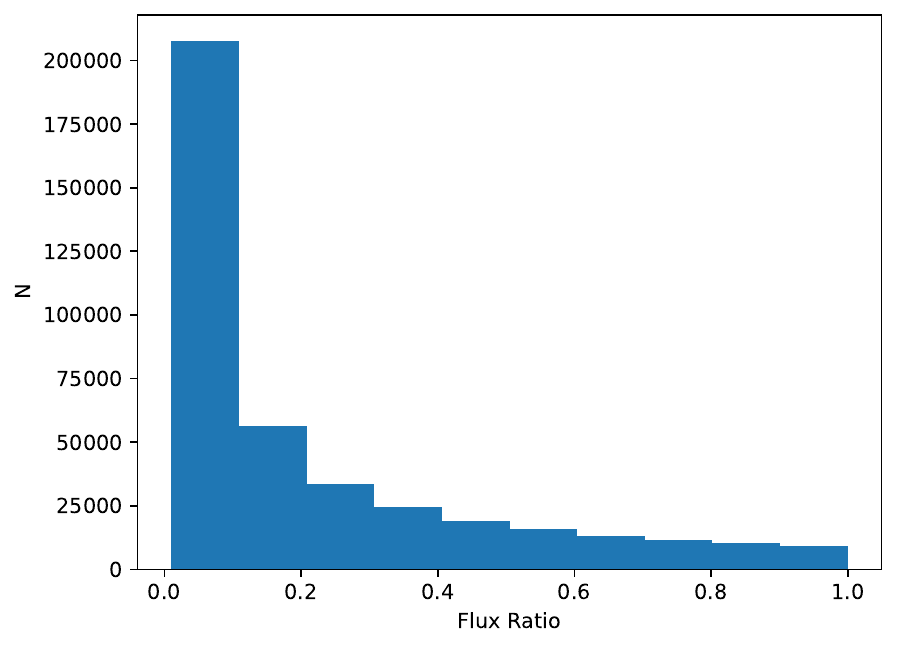}
\caption{Histogram showing the distribution of flux ratios for the simulated DESI spectra. Due to the logarithmic conversion between magnitudes and fluxes, there is a tendency to generate lower flux ratios. \label{fig:fr}}
\end{figure}

A model is then randomly selected each time a simulated spectrum is generated, with a random phase in the range -10 to +30 days relative to maximum light (bolometric). One can then also manually choose a magnitude difference between the galaxy and the SN, which allows for variation in the flux ratio in the final simulated spectra: a flux ratio of 1.0 means that there is only SN light and a flux ratio of 0.0 means only galaxy light is present. For the purposes of these simulations, the magnitudes were drawn uniformly from a magnitude range between 0 and 5; however since the conversion from magnitude to flux is logarithmic, the majority of the flux ratios are $< 0.2$. A histogram of the full distribution of flux ratios applied to the simulated data can be seen in Fig. \ref{fig:fr}. Next a real DESI exposure is chosen from the main survey, importantly including the metadata corresponding to the observing conditions for that specific exposure. This allows for the SN model to be simulated as if it had been observed in the same way as the selected DESI spectrum. The newly simulated SN spectrum is then resampled to the correct wavelength range and added to the galaxy flux from the DESI spectrum, thus creating a simulated SN spectrum as if it had been observed by DESI. 

By following this process we generated a total of 400525 spectra, which was the total number of spectra containing both galaxy and SN light using the aforementioned process to generate $\sim$600000 total spectra (including spectra of just galaxies). Each class was roughly equally represented with 100205 Type Ia, 99982 Type Ib, 100117 Type Ic, and 100221 Type II spectra respectively. From this, the spectra were split into training, validation, and testing sets of size equal to 72\%, 18\%, and 10\% of the total spectra, following typical values used in machine learning (though with a desire for a greater proportion of training data compared to the first version of \STag \citep{Davison2022} in order to maximise the possible variety in spectra trained on). We also decided that the training and validation sets would only use spectra with a flux ratio $> 0.2$ as those that do not satisfy this requirement have such little supernova light that it is highly unlikely \STag would be able to classify these; as such, removing them gives our ANN a better chance of learning from the data.

\subsection{Real DESI Spectra}
Whilst the entirety of training and testing of \STagII was done using the simulated DESI spectra (we note here that simulated DESI spectra refers to the combination of real DESI spectra combined with model SN light, as described in Section \ref{sec:sim}), we also wanted to test its capabilities with actual SN spectra observed by DESI. There are not many examples (relative to the total number of DESI spectra), but there have been some fortuitous cases where DESI observed a galaxy that was host to a SN. Such SNe can be identified by comparing the RA and DEC from DESI with that given for SNe in the \textit{Transient Name Server} (TNS)\footnote{\url{https://www.wis-tns.org}}, then further refining the selections by choosing only DESI observations that occur within some suitable time frame around the reported TNS discovery date. For our case specifically we chose observations that were made $\pm3$ weeks of the discovery date, giving us a far greater chance of the DESI spectra containing SN light.

Using these parameters we end up with a total of 8 DESI observations that match with a TNS report, covering a mean Julian date (MJD) range from 59319.4 to 59550.2, which we show in Table \ref{tab:tns}. One can quickly see an issue that has arisen, in that the reported redshift for some of the objects are different between DESI and TNS. The redshift calculated in DESI is done so by an algorithm called \texttt{Redrock} \citep{RedrockBailey2024}, which finds the best-fitting template in order to estimate the redshift. However, since DESI is not a SN survey these templates do not contain any examples of SN spectra, and as such any attempt by \texttt{Redrock} to estimate the redshift of a SN spectrum can result in an erroneous redshift. A good example of this is seen for the \texttt{Redrock} redshift of SN 2021zfs, which is significantly different to the redshift reported by TNS. Since an incorrect classification will result from the wrong redshift being used, it is important to ensure that the \texttt{Redrock} redshift is indeed reliable.

We were also fortunate to be working on the improvements to \STag during the time when SN 2023ixf was first discovered. Since the decision was made by the DESI collaboration to spend time observing this SN, we were able to obtain these spectra and use \STag to classify them. A brief description of SN 2023ixf and the results of the classification can be found in Appendix \ref{app:ixf}.

\begin{sidewaystable}
\caption{The type, position, time, and redshift data for the 8 SN found to be matching between DESI and TNS used in this paper. \label{tab:tns}}
\movetableright=-0.4in
\begin{tabular}{ccccccccc}
\hline
\hline
Name & DESI Target ID & TNS & RA & DEC & DESI Observation & TNS Discovery & z & z\\ 
 &  & Classification &  &  & (YYYY-MM-DD) & (YYYY-MM-DD) & (DESI) & (TNS)\\
\hline
SN 2021acgl & 39628467658035349 & Ia & 00:20:26.685 & +29:26:57.57 & 2021-10-15 & 2021-10-25 & 0.091 & 0.097\\
SN 2021aexj & 39628007786155359 & Ia & 01:24:54.635 & +09:12:32.00 & 2021-12-01 & 2021-11-19 & 0.020 & 0.047\\
SN 2021ihf & 39627818585294438 & Ia-pec & 14:32:14.661 & +01:20:14.29 & 2021-04-14 & 2021-04-03 & 0.14 & 0.14\\
SN 2021qtc & 39633200862986949 & Ia & 17:41:58.317 & +47:06:16.19 & 2021-06-28 & 2021-06-21 & 0.073 & 0.081\\
SN 2021ses & 39633425195336188 & Ia & 17:22:41.617 & +63:11:10.79 & 2021-06-29 & 2021-07-05 & 0.081 & 0.075\\
SN 2021sxf & 39633353015560190 & Ia & 17:10:10.767 & +57:17:43.63 & 2021-06-29 & 2021-07-08 & 0.081 & 0.089\\
SN 2021tdl & 39628401761323398 & II & 16:42:01.550 & +26:21:52.81 & 2021-07-05 & 2021-07-10 & 0.046 & 0.046\\
SN 2021zfs & 39627916874618396 & Ia & 21:32:29.830 & +05:20:46.72 & 2021-09-21 & 2021-09-21 & 0.68 & 0.020\\
\hline
\end{tabular}

\tablerefs{The discovery date, RA, DEC \citep{2021TNSTR3964....1D,2021TNSTR3659....1F,2021TNSTR3265....1T,2021TNSTR2413....1T,2021TNSTR2385....1F,2021TNSTR2318....1M,2021TNSTR2141....1F,2021TNSTR1052....1M}, and type \citep{2021TNSCR4020....1S,2021TNSCR3766....1S,2021TNSCR3273....1B,2021TNSCR2659....1P,2021TNSCR2581....1S,2021TNSCR2421....1S,2021TNSCR2204....1S,2021TNSCR1646....1V} are sourced from TNS. The redshifts are also taken from TNS and are from the SNe classifications reported there.}
\tablecomments{All DESI observations were found by searching a $\pm$3 week window around the reported TNS discovery date.}
\end{sidewaystable}

\section{Methodology}\label{sec:meth}
\subsection{STag}
What follows is a brief description of how \STag works, for a more detailed overview of the architecture the reader is directed to our previous paper \citep{Davison2022}.

The tag probabilities are calculated using logistic regression, a function that rapidly transitions between 0 and 1; where the probability for the tag of a particular spectral feature is given by Equation \ref{eq:lr} \citep{Cramer2005}:

\begin{equation}\label{eq:lr}
    \sigma_k(z) = \frac{1}{1 + e^{-z}},
\end{equation}

\noindent and so for a given supernova, the tag probabilities can be expressed as a vector of \textit{n} total tags:

\begin{equation}
    \vec{\sigma} = (\sigma_0,\sigma_1,...,\sigma_n).
\end{equation}

\noindent In Eq. \ref{eq:lr}, \textit{z} is as seen in Equation \ref{eq:z}:

\begin{equation}\label{eq:z}
    z = \beta_0 + \sum^{N}_{i = 1} \beta_i x_i,
\end{equation}

\noindent where $\beta_0$ is a normalisation constant, and $\beta_i$ is the weight of the flux value $x_i$. The beta values (excluding $\beta_0$) form a visual representation of the actual shape of the spectral feature in question, and as such when Equation \ref{eq:lr} is close to 1, the spectrum has a high probability of showing said feature (and vice versa if Equation \ref{eq:lr} is close to 0). The beta values were determined using a combination of a loss function (which should be minimised in order to obtain the best values) as seen in Eq. \ref{eq:ce}:

\begin{equation}\label{eq:ce}
    J = - \frac{1}{N} \sum^N_{i=1} [Y_i\ln{\sigma(z)_i} + (1-Y_i)\ln{(1-\sigma(z)_i)}],
\end{equation}

\noindent where \textit{J} is the loss, \textit{N} is the number of objects to be tagged, and $Y_i$ is either equal to 1 or 0 depending on if the object has the tag or not respectively. We also employ stochastic gradient descent which uses a subset of the data to change the gradients \citep{Bottou2010} in order to optimise accuracy vs. running time.

Once a spectrum has all the tag probabilities calculated, these are then passed to a simple feed-forward ANN \citep{Rumelhart1986}. This consists of an input layer which takes the tag probabilities, which is then fully-connected to a single hidden layer, which itself connects to the output layer. The output layer determines the best suited class using softmax regression, which is given by Equation \ref{eq:soft}:

\begin{equation}\label{eq:soft}
    s(y)_i = \frac{e^{y_i}}{\sum\nolimits_{j=1}^N e^{y_j}},
\end{equation}

\noindent where $y$ is a vector consisting of the weights from the output layer of the ANN. The output is a vector of class probabilities, which all sum up to 1. The ANN learns that different classes of SN have distinct tag probabilities and so is able to accurately classify spectra.

\subsection{Tags}
\begin{figure}[htp]
\centering
  \begin{subfigure}[b]{0.75\textwidth}
    \centering
    \includegraphics[width=1\linewidth]{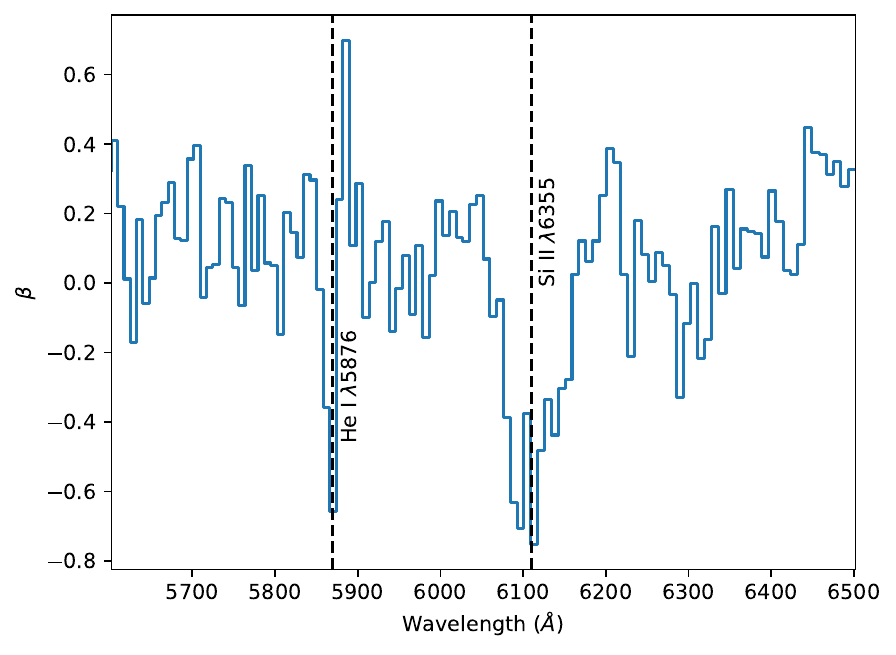} 
    \caption{\STag \SiII\ $\lambda$6355 tag} 
    \label{fig:tagold} 
  \end{subfigure} 

  \begin{subfigure}[b]{0.75\textwidth}
    \centering
    \includegraphics[width=1\linewidth]{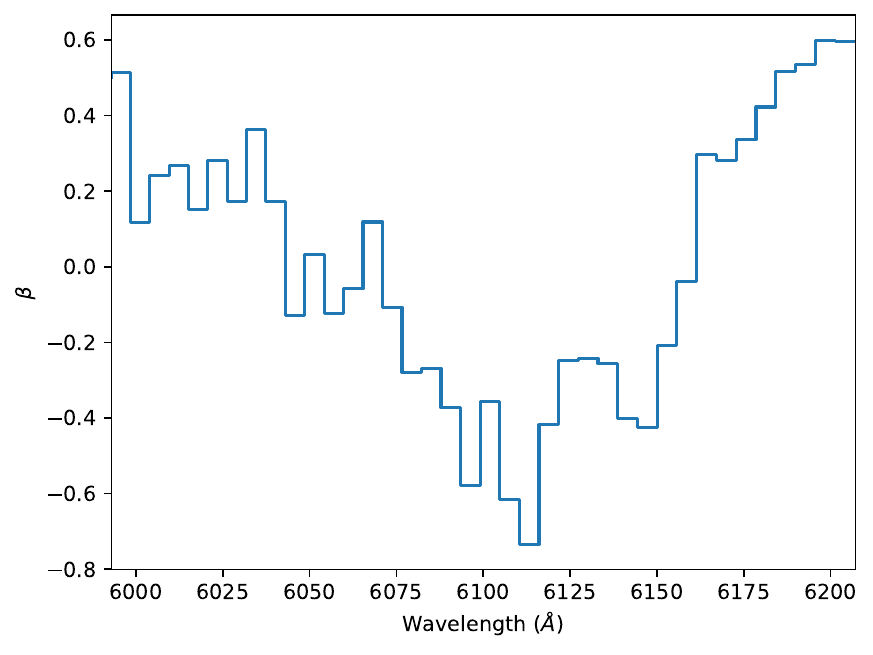} 
    \caption{\STagII \SiII\ $\lambda$6355 tag} 
    \label{fig:tagnew} 
  \end{subfigure}
  \caption{Comparison of the original $\beta$ values for the \SiII\ $\lambda$6355 tag (originally labelled as \SiII\ $\lambda$6150 in \citealt{Davison2022}), which can be seen to also possibly include the absorption feature associated with \HeI\ $\lambda$5876, and the $\beta$ values for the \SiII\ $\lambda$6355 tag used in \STagII, which now only contains the relevant feature to avoid potential contamination from other features.}
\label{fig:tagcomp}
\end{figure}

With the introduction of a new dataset, it was necessary to also revisit the tags and recreate them making use of the new data. Some of the original tags had long wavelength ranges that potentially included other features (see Figure \ref{fig:tagold}) and in the case of the hydrogen tag, utilised the whole spectrum. As such it was unclear exactly what it was using to determine the presence of hydrogen. This led to the recreation of all the original tags with stricter wavelength limits (see Figure \ref{fig:tagnew}), as well as the introduction of tags for individual hydrogen features.

With the new tags created, the neural network was repeatedly trained and tested with different combinations of included tags in order to determine which tags were necessary for the classification of a SN and which were superfluous. The result of this extensive training and testing was that the \HeI\ $\lambda$6678 tag (which is characterised by a doublet feature) was causing a lot of confusion between classifications and as such was removed. Specifically, since this feature was very seldom seen in spectra outside of Type IIb SNe, if it had even a $<10\%$ tag probability it was enough to reduce the softmax probability of the predicted class. This had the knock-on effect of also necessitating the removal the Type IIb class from \STag's functionality as there was no longer a suitable spectral feature to differentiate it from the over-arching Type II class. Another change that was borne out of these tests was that it was decided that an increase in the number of wavelength points of the spectra would help account for smaller changes in the tags. The original value for the number of wavelength points was chosen to be $R = 1024$ as this was the default setting used in \texttt{DASH}, selected so as to be able to distinguish between broad and narrow features whilst maintaining the desired speed of the classifier \citep{Muthukrishna2019}. As such, to accommodate for a greater capability of catching the more subtle changes of a spectral feature whilst also avoiding introducing too much noise or making \STag drastically more computationally expensive, the number of points that make up the spectra was increased from 1024 to 1500. This value was chosen after trialling different numbers of wavelength points up to 2048, whereupon it was ultimately decided that the value of 1500 allowed for the best balance of finer details and computational cost. 

The \FeII\ tag was also further reduced, instead of being one tag that encompassed three suspected features it is now a tag specifically for the \FeII\ $\lambda$5170 feature as it was found to be the most frequent and readily identifiable. It was also decided that the distinction between an emission, absorption, and P-Cygni line profile for the \HeI\ $\lambda$5876 feature was unnecessary and likely causing more confusion than clarity when reporting on tag probabilities (it was not uncommon for spectra to have high probabilities for both absorption and emission tags; \citealt{Davison2022}). As a result, only the absorption feature is now explicitly tagged for and results in a clear identifier of the \HeI\ $\lambda$5876 absorption feature.

To further strengthen the classifications, as well as to help the neural network to make the correct classification in the first place, a number of new tags were also introduced. These consist of a tag for \SiII\ $\lambda$4000, H$\alpha$ and narrow H$\alpha$ $\lambda$6563 (the latter henceforth referred to as n-H$\alpha$, see Appendix \ref{app:ixf} for a more detailed description of these two hydrogen lines), and H$\beta$ $\lambda$4861. Whilst not requiring a new tag specifically, the spectral features of \SiII\ $\lambda$5876 and \NaI\ D $\lambda$5876 both make use of the tag for \HeI\ at the same wavelength. All three of these features overlap at approximately the same wavelength so even with the more restricted ranges of the tags, it would still be impossible to create separate tags for each of these. We therefore make use of the fact that the \HeI\ $\lambda$5876 tag is now only for an absorption feature and that both \SiII\ $\lambda$5876 and \NaI\ D $\lambda$5876 are also absorption features to effectively create a multi-purpose tag. 

The final set of tags included in \STagII are H$\alpha$, n-H$\alpha$, and H$\beta$ for Type II SNe, \CaII\ H\&K, \SiII\ $\lambda$4000, \SiII\ $\lambda$6355, and \SII\ for Type Ia SNe, \HeI\ $\lambda$5876 for Type Ib SNe, and \FeII\ $\lambda$5170 as a general tag. The differences between the tags included in the two different versions of \STag can be found in Table \ref{tab:tags_comp}.

\begin{table}[tbp]
\begin{center}
\caption{Table comparing the tags present in the two different versions of \STag. \label{tab:tags_comp}}
\begin{tabular}{ccccc}
\hline
\hline
Tag & \STag & \STagII & Wavelength Range (\AA) & Central Wavelength (\AA)\\
\hline
\CaII\ H\&K & $\checkmark$ & $\checkmark$ & 3652 - 3892 & 3770\\
\SiII\ $\lambda$4000 & $\times$ & $\checkmark$ & 3892 - 4076 & 3983\\
H$\beta$ & $\times$ & $\checkmark$ & 4588 - 4752 & 4669\\
\FeII\ $\lambda$5170 & $\times$ & $\checkmark$ & 5141 - 5256 & 5198\\
\SII\ & $\checkmark$ & $\checkmark$ & 5188 - 5607 & 5393\\
\HeI\ $\lambda$5876 (Emission) & $\checkmark$ & $\times$ & -- & --\\
\HeI\ $\lambda$5876 (Absorption) & $\checkmark$ & $\checkmark$ & 5520 - 5818 & 5667\\
\HeI\ $\lambda$5876 (P Cygni) & $\checkmark$ & $\times$ & -- & --\\
\SiII\ $\lambda$6355 & $\checkmark$ & $\checkmark$ & 5993 - 6207 & 6099\\
H$\alpha$ & $\times$ & $\checkmark$ & 6358 - 6702 & 6528\\
H$\alpha$ (Narrow) & $\times$ & $\checkmark$ & 6507 - 6622 & 6564\\
\HeI\ $\lambda$6678 & $\checkmark$ & $\times$ & -- & --\\
\hline
\end{tabular}
\end{center}
\tablecomments{The \FeII\ tag was present in the first version of \STag as a combination of three lines rather than just the one in \STagII. It should also be noted that there was a hydrogen tag in the first version of \STag, however it used the full spectrum and so wasn't specific to any one feature \citep{Davison2022}. The wavelength range quoted here is the range across which the tag is applied for the respective feature.}
\end{table}

\subsection{Equivalent Width}
The equivalent width (\textit{W}) of a line is defined as in Equation \ref{eq:eqw}:

\begin{equation}\label{eq:eqw}
    W = \int \frac{F_c - F_l}{F_c} d\lambda,
\end{equation}

\noindent where \textit{$F_c$} is the intensity of the continuum and \textit{$F_l$} is the intensity of the spectral line. For an absorption line the equivalent width is positive, whilst for an emission line it is negative. The equivalent widths in this paper were calculated using the \texttt{specutils} package \citep{nicholasearl2023}; which is itself a package of \texttt{Astropy} \citep{astropy:2013,astropy:2018}. All tags used in \STagII also have an equivalent width calculated for the same feature using the same wavelength range as for the tag, with the exception of the sulphur line as this is a \lq W\rq\ shaped feature and so it is less obvious what the equivalent width of this feature would correspond to. 

In order to see if there was extra information that could be used between the probability of a given tag and the equivalent width of the associated spectral feature, we also trialled including a nonlinear combination of these two values by passing the product of each pairing as separate inputs to the neural network. Ultimately it was determined that such nonlinear combinations offered no extra information for which the neural network could use to improve the classifications, with the exception of the Si $\lambda$4000 feature. It is unclear why this feature, and this alone, makes a difference on the accuracy of the classifications, however its inclusion does lead to greater accuracy and so it continues to be utilised.

The inclusion of equivalent widths, and the changes to the tags, meant that the architecture of \STag had to be changed. \STag now consists of an input layer of 18 nodes, followed by three fully connected layers of 48 nodes each, then two layers of 96 nodes each, which then connects to two more layers of 48 nodes each, and finally a softmax output layer consisting of 4 nodes. Each of the hidden layers make use of the Rectified Linear Unit (ReLU) activation function \citep{nair2010rectified}. This neural network was built using a combination of both \texttt{keras} \citep{chollet2015keras} and \texttt{TensorFlow} \citep{tensorflow2015-whitepaper} and has the architecture it does after extensive testing to find the best-performing number of hidden layers and nodes.

\subsection{rlap Score}
In order to give us a further measure of confidence in a classification beyond just the associated softmax probability, we also made use of what is known as the \textit{r}lap score, as devised by \citep{Blondin2007}. A value of 0 would mean that there is no correlation between two spectra being considered, whilst it is typically accepted that a value above $\sim$5-6 suggests a reasonably strong correlation \citep{Blondin2007,Muthukrishna2019}. This parameter is built off of first cross-correlating the spectrum of the SN with that of a template spectrum that is at $z=0$:

\begin{equation}\label{eq:cc}
    c(n) = s(n) \star t(n) = t(n) \star t(n-\delta) + a(n).
\end{equation}

\noindent Here $c(n)$ is the cross-correlation, $s(n)$ is the SN spectrum, $t(n)$ is the template spectrum, $t(n-\delta)$ is the template spectrum shifted by some amount $\delta$ in logarithmic wavelength space, and $a(n)$ is a function that distorts the peak of the correlation function \citep{Blondin2007}. 

\begin{figure}[tbp]
\centering
\includegraphics[]{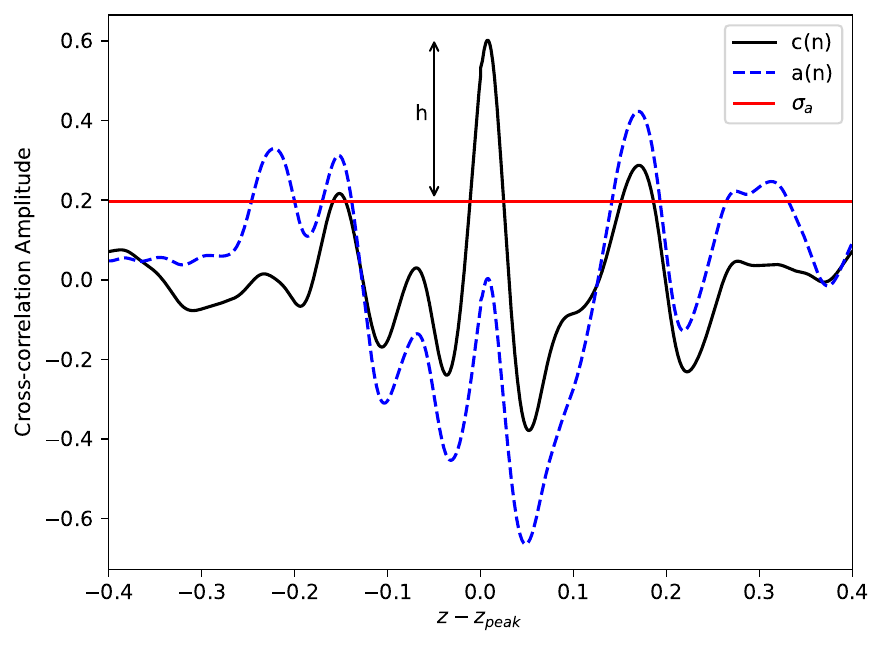}
\caption{The cross-correlation function of a real DESI spectrum with a Type Ia template spectrum. This is a functional recreation of similar plots found in \citep{Blondin2007} and \citep{Muthukrishna2019}, found in Figures 5 and 3 respectively. \label{fig:cc}}
\end{figure}

The \textit{r}lap score is the product of two values: \textit{r}, which is the cross-correlation height-noise ratio \citep{Tonry1979} which is given by Equation \ref{eq:r}, and lap, which is a measure of the overlap in wavelength space of the two spectra being cross-correlated and is defined by Equation \ref{eq:l} \citep{Blondin2007}: 

\begin{equation}\label{eq:r}
    r = \frac{h}{\sqrt{2}\sigma_a},
\end{equation}

\begin{equation}\label{eq:l}
    \mathrm{lap} = \ln\frac{\lambda_{1}}{\lambda_{0}},
\end{equation}

\noindent where \textit{h} is the height of the peak of the cross-correlation compared to the rms of the anti-symmetric component, given by $\sigma_a$. $\lambda_1$ and $\lambda_0$ are the maximum and minimum wavelength the input spectra overlaps with the comparison spectra; see Figure \ref{fig:cc} for an example of how to interpret Equation \ref{eq:r} visually. Note that we are able to use \textit{z} here since we are working in log wavelength space and so a shift in this space is equivalent to a linear shift in $1+z$ \citep{Blondin2007}.

By using both the value of \textit{r} and of lap one has a way of quantifying how similar a given spectrum is to a template spectrum and so is an extra level of certainty that our classification is accurate. To streamline the process, we assume the softmax classification is the best estimation of the class and so only cross-correlate a given spectrum with templates of the same type, but with differing phase. This now gives \STag the added functionality of estimating the phase of a SN, though we stress this is not the primary function of \STag and the range of phases is used primarily to take care of the fact that SN spectra change with time and so prevent a dilution of the \textit{r}lap value due to comparison with many spectra of the correct type but wrong phases.

We make use of 5 different phase bins, which are based on the phase range used to generate the DESI template spectra (and as such the DESI simulated spectra). The first bin encompasses all negative phase spectra (with respect to maximum light), the next 3 bins are separated by 7 days each, and the final bin is for all phases greater than 21 days since maximum light, up to 30 days.

The cross-correlation process was adapted from that used in \texttt{DASH} \citep{Muthukrishna2019}, with minor adjustments made to fit the pipeline of \STag. These changes included, but are not limited to, adjusting the wavelength points to consider (since we increased the resolution of \STag from 1024 to 1500 points) and a change to account for the reading of the different template spectra being used, but functionally it works the same as in \texttt{DASH}. 

\subsection{Redshift Checking}\label{sec:zcheck}
\begin{figure}[tbp]
\centering
\includegraphics[]{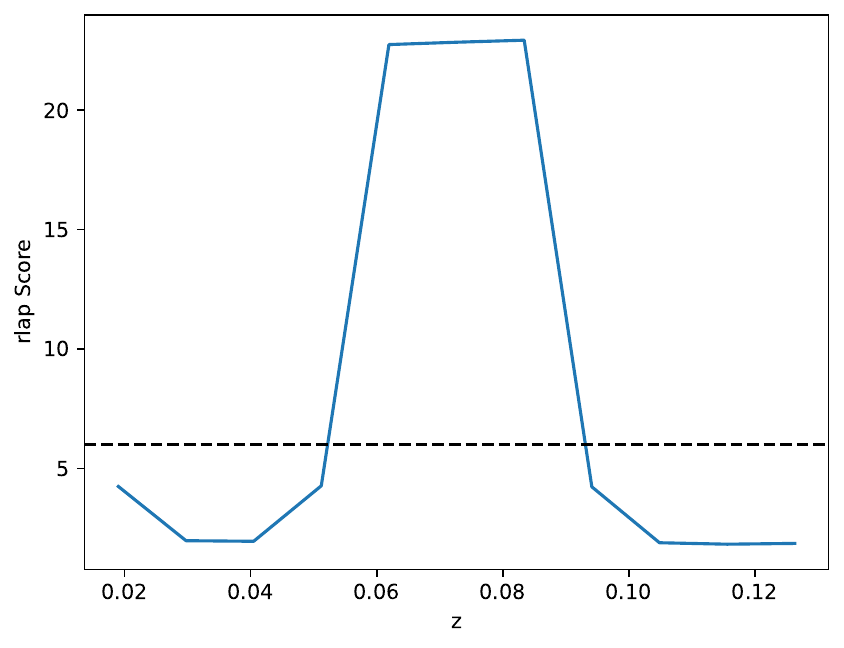}
\caption{The \textit{r}lap score for the spectra of SN 2021qtc when pre-processed at different redshifts. The \textit{r}lap score drops quickly outside $\sim$1\% of the true redshift value, meaning only when the redshift is accurate will the \textit{r}lap score be able to pass the cutoff point of 6, indicated by the horizontal black dashed line. \label{fig:rlapz}}
\end{figure}

\STag is highly dependent on accurate redshifts for it to be able to classify a spectra correctly, as such it is important that any redshifts being supplied are correct. However, \STag is not designed to be something that can estimate a redshift. As such, we are not interested in producing a comprehensive framework for calculating the redshift of a SN, though we note that \STag could be modified to do this by using the \textit{r}lap score. Instead, we adopt the philosophy that if a redshift is significantly inaccurate, we accept that \STag will simply be unable to return a satisfactory classification. We decided that should a spectrum initially fail the \textit{r}lap criteria, this may be caused by an incorrect redshift and so we can then perform a check for a degree of inaccuracy. In the case of a relatively small inaccuracy it is worth being able to potentially use a more accurate redshift as this will lead to a better classification, as well as the possibility of a spectrum passing the \textit{r}lap cut that it would have previously failed. The general principle we used was to check a series of redshifts within 10\% of 1+z (where $z$ is the redshift returned by \texttt{Redrock}), resulting in a check of 5\% either side of the given redshift. We chose to look at just 10 redshifts in the proposed range (balancing speed with enough redshift values to properly explore values around the given redshift), which means that each redshift is roughly 1\% from its surrounding redshift values. What we found was that a spectra would pass the final \textit{r}lap cut if the redshift was within 1\% of the true redshift, whilst the \textit{r}lap value would quickly fall off outside of this limit (see Figure \ref{fig:rlapz}). Therefore, should none of the 10 alternative redshifts pass the \textit{r}lap criteria then \STag will not return a classification and indicate to the user that it is possibly being caused by an incorrect redshift value.

\section{Results}\label{sec:res}
\subsection{Testing Data Comparison}
\begin{figure}[tbp]
\centering
\includegraphics[]{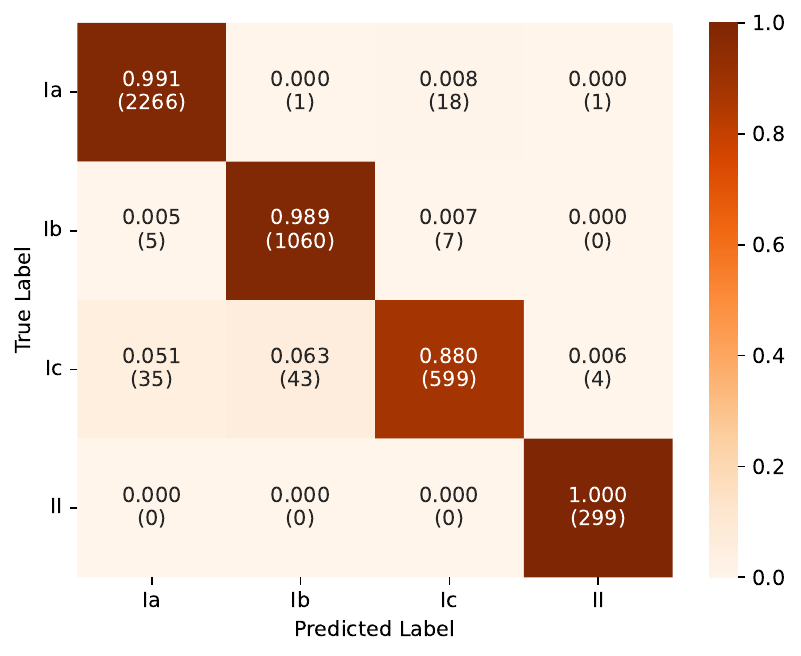}
\caption{The normalised confusion matrix of the DESI simulated spectra set aside for testing \STagII, which compares the estimated classification to the true class (as determined by which SN model was used). The numbers in brackets below the normalised fraction that were correctly classified correspond to the number of spectra that were classified as that class by \STagII. \label{fig:conf}}
\end{figure}

We made use of the DESI simulated data (as detailed in Section \ref{sec:sim}) that had been set aside for use as testing data, which was a total of 40053 spectra (10\% of the total spectra). All of these spectra were classified by \STagII, before having the best \textit{r}lap score calculated for each one. After applying our cut of only accepting classifications with an \textit{r}lap score $>$ 6, the final number of spectra was 4338. Whilst we note that this is a sizeable reduction in the number of spectra (approximately 90\% of the spectra in the testing set do not pass our \textit{r}lap cut), it is important to remember that many of the simulated spectra have very low flux ratios ($< 0.2$) and so will fail the \textit{r}lap cut due to having very little supernova light. There is no clear flux ratio threshold correlating with a spectra passing the \textit{r}lap cut, and whilst higher values of the flux ratio do lead to better classifications and the degree by which the \textit{r}lap cut is passed, any spectra with a flux ratio $>0.2$ can be successfully classified by \STagII. We present the normalised confusion matrix for the classifications of these spectra in Figure \ref{fig:conf}.

We note a very high accuracy for all classes, with both Type Ia and Ib SN having a 99\% accuracy. The classification success of Type Ic SN is lower at 88\%, though this is likely due to the lack of a clear, defining characteristic of the Type Ic class. We also consider the 100\% success rate for Type II SNe to be misleading, as this is likely a result of the fact that when creating the simulated data for each type of SN, there were fewer unique/distinct models of Type II SN compared to the other types and so \STagII is learning the specific details of the SN models used, rather than of the Type II archetype overall. Still, assuming that the models used are truly representative of the full population of Type II SNe we expect that \STagII is capable of classifying this type reliably (though, as a counter example, we refer the reader to Appendix \ref{app:ixf}).

\subsection{TNS Classifications}
We classified the 8 SN listed in Table \ref{tab:tns}, including the functionality of \STag to report back whether a redshift had been changed from the input value, and if so what the new redshift was. We find that we are able to correctly classify 2 of the SN after the \textit{r}lap cut is applied, with Table \ref{tab:tns_class} (see Appendix \ref{app:tns}) showing these along with any significant tag probabilities. This lack of completeness is a result of the mismatch between DESI observation date and the TNS discovery date, with the observation date often being before the reported discovery, which results in a lack of supernova light in the DESI spectra. We now discuss each of the 8 cases in detail, with explanations for a missing classification where applicable.

\subsubsection{SN 2021acgl}
This supernova has a reported classification as a Type Ia and since the \texttt{Redrock} redshift is very close to that from TNS, it is likely that \STag was unable to make a classification due to a combination of a low probability for the \SiII\ $\lambda$6355 tag and the fact that the DESI observation took place 10 days prior to the TNS discovery date, which itself was before maximum light \citep{LASAIR}. One can see from Figure \ref{fig:acgl} that there is practically no clear supernova light in the DESI spectra, owing to this earlier observation date.

\subsubsection{SN 2021aexj}
We find that \STag is able to correctly classify this SN as a Type Ia, reflected in the tag probabilities that are present (most importantly, 100\% for \SiII\ $\lambda$6355). There has been a change of the redshift made by \STag, going from the \texttt{Redrock} redshift of 0.020 to 0.051. We note that the new redshift is closer to that of the TNS redshift and so are confident the redshift checking method is working satisfactorily; a comparison of the DESI spectra to one on TNS can be seen in Figure \ref{fig:aexj}. The phase bin corresponding to the highest \textit{r}lap score is  $14 \leq t \leq 21$, which is possibly older by around 7 days than the age as determined from the light curve \citep{LASAIR}.

\subsubsection{SN 2021ihf}
Since a Ia-pec SN is defined by the fact that it has an unusual or unique spectra for a Type Ia \citep{Blondin2007}, it is unlikely to look similar to any of the models used to produce the simulated spectra or for the cross-correlation process when calculating the \textit{r}lap (the spectra of this SN can be seen in Figure \ref{fig:ihf}). The classification report for this supernova notes it is similar to SN 2000cx \citep{2021TNSCR1646....1V}, which is characterised by the presence of weaker \SiII\ features in early spectra \citep{Li2001}. This is reflected by the fact that the tags for the \SiII\ features are quite low and as such it is unlikely \STag would be able to classify such a SN.

\subsubsection{SN 2021qtc}
This observation took place just over a week after the TNS discovery date so there is definitely SN light present in this spectrum, reflected by the 100\% probability of the \SiII\ $\lambda$6355 tag which can be seen clearly in Figure \ref{fig:qtc}. The predicted phase of $0 < t \leq 7$ lines up well with the time of maximum light as well \citep{LASAIR}, causing the clear spectral features.

\subsubsection{SN 2021ses}
The presence of H$\alpha$, H$\beta$, and \HeI\ $\lambda$5876 with high probabilities is interesting as this would suggest a Type II SN instead of the expected Type Ia. However, inspecting the DESI spectra seen in Figure \ref{fig:ses} it is clear that there is no H$\alpha$ and instead an absorption feature appears to be mistakenly triggering the tag for broad H$\alpha$ (which does include an absorption component). The DESI observation date is also 15 days before maximum light \citep{LASAIR} and so there is likely little to no SN light in the spectrum.

\subsubsection{SN 2021sxf}
This spectrum has a deficit of spectral features, with only \HeI\ $\lambda$5876 having a high probability (97\%) which would likely lead to a Ib classification, if anything, and not the Type Ia we would expect from TNS. From the DESI spectrum seen in Figure \ref{fig:sxf} one can see that there is possibly a very broad yet very shallow absorption feature roughly where we would expect \SiII\ $\lambda$6355 to be. Since this spectrum was observed 9 days before the TNS discovery date and approximately 20 days before maximum light \citep{LASAIR}, it is likely the DESI spectra is simply too early to properly see the necessary features.

\subsubsection{SN 2021tdl}
This was classified as a Type II SN on TNS and there is a clear H$\alpha$ emission in the TNS spectra, whereas there is no such feature in the DESI spectrum (see Figure \ref{fig:tdl}). Much like many of the SN we were unable to classify, SN 2021tdl was observed by DESI before it was reported on TNS and so does not contain much, if any, SN light.

\subsubsection{SN 2021zfs}
There are multiple factors which contribute to \STag being unable to make a classification for this spectrum. Firstly, the \texttt{Redrock} redshift is significantly different to the TNS redshift, and is well beyond the 10\% check we do in redshift. As such the spectral features are massively offset, to the degree that a significant portion of the spectrum is no longer within the wavelength range being considered and so the tag probabilities do not correspond to the actual features within the spectrum (see Figure \ref{fig:zfs}). With the TNS redshift used, we are able to correctly classify this supernova as a Type Ia. However, since this is not information we would have access to when using \STag as part of the pipeline we do not report this successful classification in Table \ref{tab:tns_class}.

\begin{figure}[ht]
  \begin{subfigure}[b]{0.48\linewidth}
    \centering
    \includegraphics[width=0.825\linewidth]{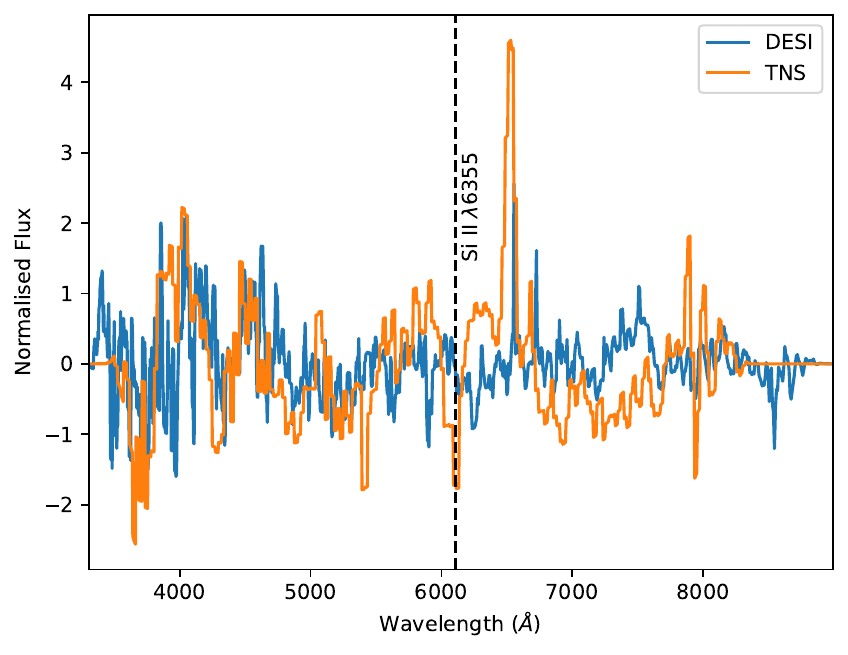} 
    \caption{SN 2021acgl ($\Delta t_{(\mathrm{DESI} - \mathrm{TNS})}$ = -20 days)} 
    \label{fig:acgl} 
  \end{subfigure} 
  \hfill  
  \begin{subfigure}[b]{0.48\linewidth}
    \centering
    \includegraphics[width=0.825\linewidth]{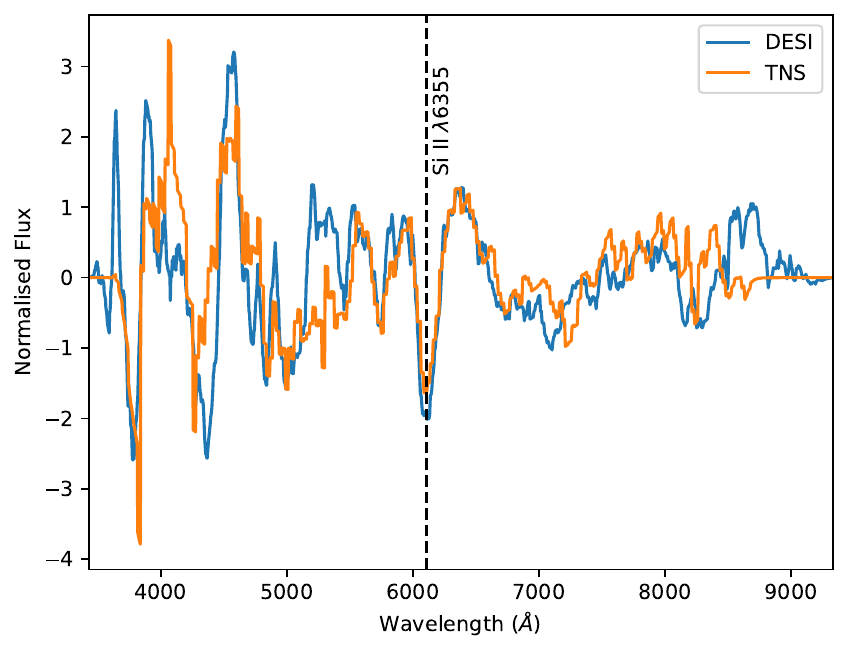} 
    \caption{SN 2021aexj ($\Delta t_{(\mathrm{DESI} - \mathrm{TNS})}$ = +7 days)} 
    \label{fig:aexj} 
  \end{subfigure} 

  \begin{subfigure}[b]{0.48\linewidth}
    \centering
    \includegraphics[width=0.825\linewidth]{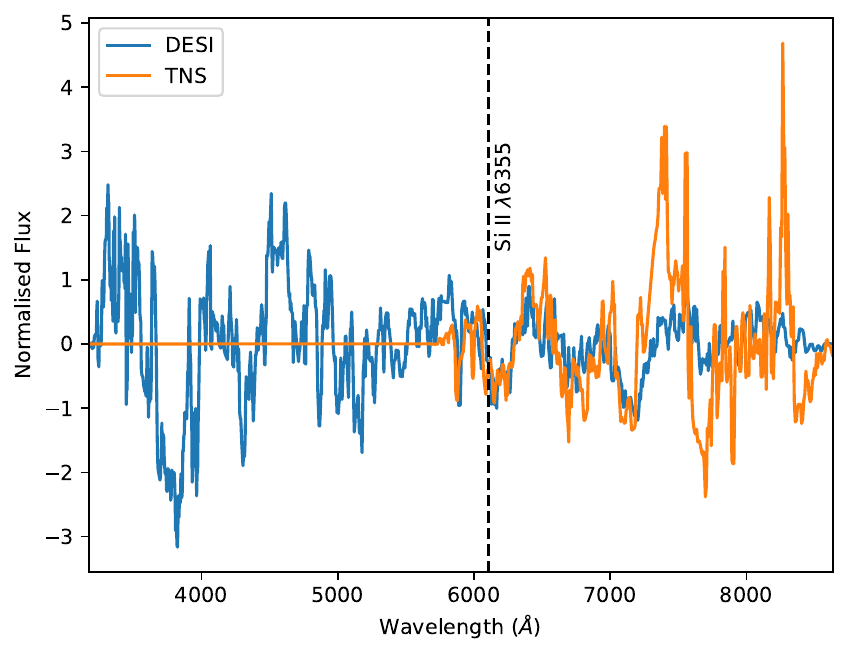} 
    \caption{SN 2021ihf ($\Delta t_{(\mathrm{DESI} - \mathrm{TNS})}$ = -10 days)} 
    \label{fig:ihf} 
  \end{subfigure} 
  \hfill
  \begin{subfigure}[b]{0.48\linewidth}
    \centering
    \includegraphics[width=0.825\linewidth]{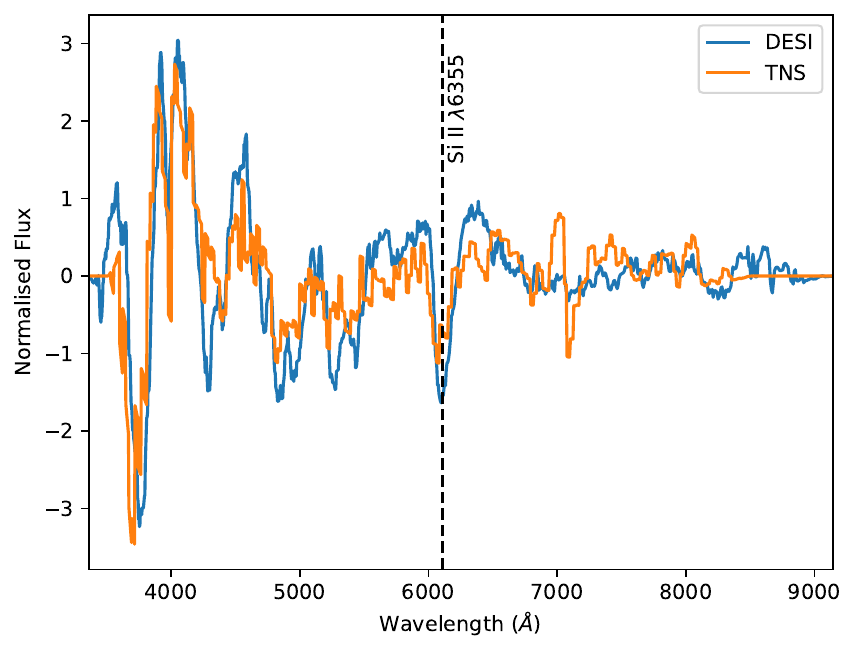} 
    \caption{SN 2021qtc ($\Delta t_{(\mathrm{DESI} - \mathrm{TNS})}$ = +3 days)} 
    \label{fig:qtc} 
  \end{subfigure} 

  \begin{subfigure}[b]{0.48\linewidth}
    \centering
    \includegraphics[width=0.825\linewidth]{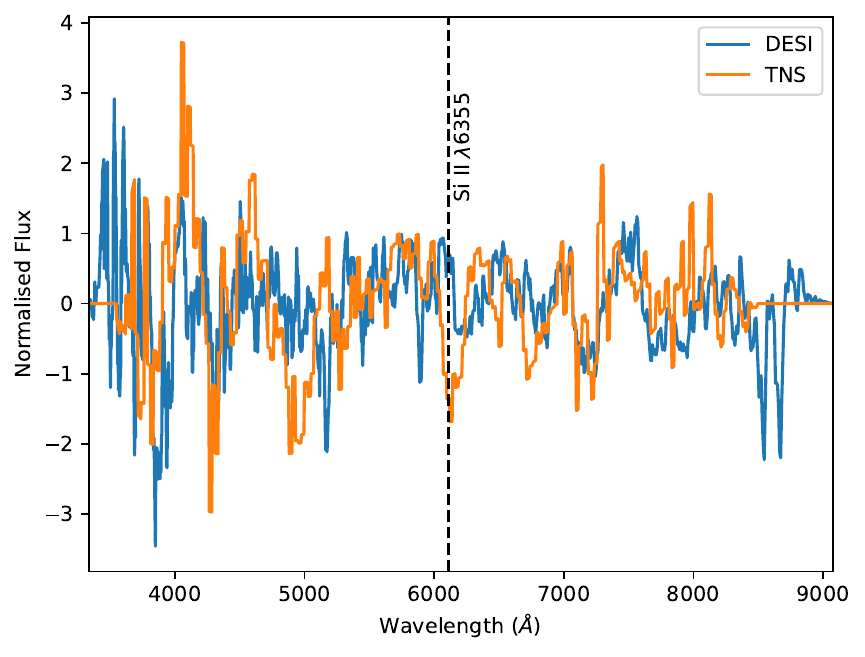} 
    \caption{SN 2021ses ($\Delta t_{(\mathrm{DESI} - \mathrm{TNS})}$ = -13 days)} 
    \label{fig:ses} 
  \end{subfigure}
  \hfill
  \begin{subfigure}[b]{0.48\linewidth}
    \centering
    \includegraphics[width=0.825\linewidth]{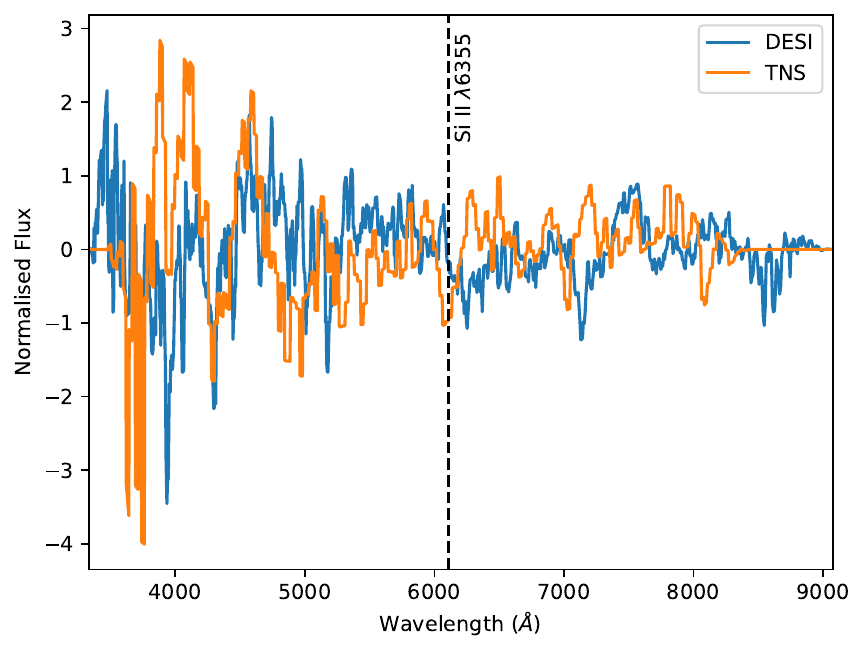} 
    \caption{SN 2021sxf ($\Delta t_{(\mathrm{DESI} - \mathrm{TNS})}$ = -28 days)} 
    \label{fig:sxf} 
  \end{subfigure} 
\end{figure}

\clearpage

\begin{figure}[ht]
\ContinuedFloat
   \begin{subfigure}[b]{0.48\linewidth}
    \centering
    \includegraphics[width=0.825\linewidth]{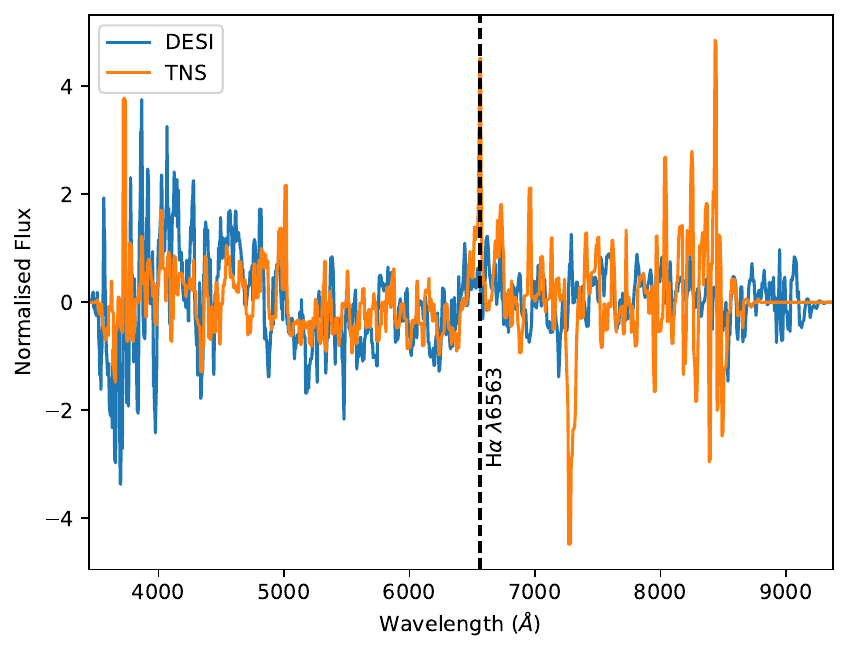} 
    \caption{SN 2021tdl ($\Delta t_{(\mathrm{DESI} - \mathrm{TNS})}$ = -28 days)} 
    \label{fig:tdl} 
  \end{subfigure}
  \hfill
  \begin{subfigure}[b]{0.48\linewidth}
    \centering
    \includegraphics[width=0.825\linewidth]{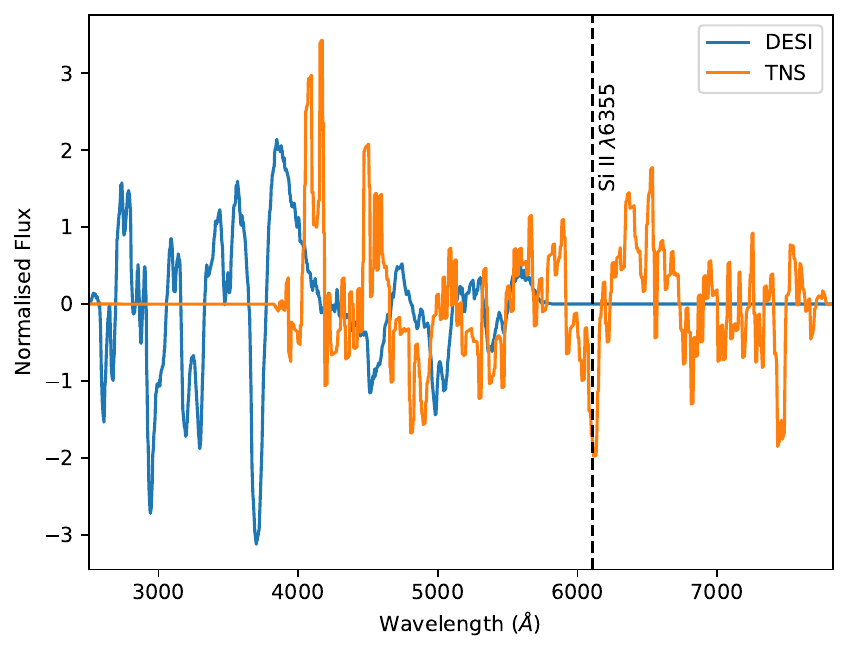} 
    \caption{SN 2021zfs ($\Delta t_{(\mathrm{DESI} - \mathrm{TNS})}$ = -1 days)} 
    \label{fig:zfs} 
  \end{subfigure} 

\caption{Comparison of DESI observed spectra and spectra of each supernova sourced from TNS, adjusted to the rest-frame wavelengths. $\Delta t_{(\mathrm{DESI} - \mathrm{TNS})}$ is the time difference between the date the DESI observation was made and when the TNS spectra \citep{2021TNSCR3766....1S,2021TNSCR4020....1S,2021TNSCR1646....1V,2021TNSCR2204....1S,2021TNSCR2421....1S,2021TNSCR2581....1S,2021TNSCR2659....1P,2021TNSCR3273....1B} was observed.}
\label{fig:compspec} 
\end{figure}

\section{Discussion}\label{sec:dis}
Unlike as with the first version of \STag, which always gave a classification regardless of whether it was accurate or not, the introduction of the \textit{r}lap score cut means that \STagII does not always return a classification. As such, it is harder to judge the importance of certain tags in making certain classifications as before. However, it is worth noting that the 2 cases that were classified from the TNS selection both have very high probabilities of \SiII\ $\lambda$6355 whilst all other cases (that were Type Ia) do not have a high probability of this feature. It is reasonable to assume that \SiII\ $\lambda$6355 is still the dominant tag for a Type Ia classification owing to it being universally present in Type Ia SNe and frequently very strong, while \SII\ does not seem to have an impact (SN 2021aexj has essentially no \SII\ feature, whilst for SN 2021qtc it is at 100\%). Furthermore, based on the fact that the Type Ia SNe that were not classified have varying tag probabilities for \CaII\ H\&K and \SiII\ $\lambda$4000, it is deemed that these are sub-dominant to \SiII\ $\lambda$6355 and are not enough on their own to trigger a Ia classification.

Whilst there are no Ib or Ic SNe in the TNS selection, it is worth noting that the \HeI\ $\lambda$5876 feature consistently has a high probability, highlighting its problematic nature as a diagnostic of purely a Ib SN. Indeed, \cite{Liu2016} suggest that for a confident identification of the feature at 5876 \AA{} as \HeI\ and not due to a different line there should either be a detection of 2 other \HeI\ lines ($\lambda$6678 and $\lambda$7065) or the \HeI\ $\lambda$5876 line should be very strong before maximum light. The issue with the former method is that the $\lambda$6678 line has significant overlap with H$\alpha$ and the $\lambda$7065 line is at a sufficiently high wavelength that noise tends to start to dominate the spectrum. The second method may be possible to implement due to the inclusion of a phase estimate from \STag, however at this time such a decision is not automated and would have to be done by manually inspecting the relevant values.

Unfortunately the only Type II SN in the TNS selection was unable to be classified, so it is also hard to draw a resolute conclusion about the dominance of the H$\alpha$, n-H$\alpha$, and H$\beta$ tags. However, it does not seem to sway the classification if H$\alpha$ is present at relatively high probabilities if there are other deterministic features present. The extra tag of \FeII\ $\lambda$5170 continues to vary wildly between different spectra and so it is assumed that, as was concluded in \cite{Davison2022}, it does not have a dominant affect on a classification result.

\subsection{\STagII vs. \texttt{DASH}}
Since \STag was originally inspired by and created to be competitive with the machine learning based spectral classifier \texttt{DASH} \citep{Muthukrishna2019}, we also used it to classify the 8 DESI spectra that have TNS classifications. We found that \texttt{DASH} was only able to correctly classify SN 2021aexj as a Type Ia SN. The other case \STag was able to correctly classify, SN 2021qtc, was incorrectly classified by \texttt{DASH} as a Type Ic. For SN 2021acgl, SN 2021ses, Sn 2021sxf, and SN 2021zfs the classifications are all Type Ic-broad, a classification result which \texttt{DASH} often considers to be unreliable as it usually results from spectra with high amounts of host light \citep{Muthukrishna2019}. Both SN 2021ihf and SN 2021tdl are incorrectly classified as a Type Ib and a Type Ia respectively. Not only does \textit{STag} manage to classify an extra spectra compared to \texttt{DASH}, by not reporting a classification result if the \textit{r}lap cut is not met, we also avoid reporting incorrect classification results.

\subsection{\STag vs. \STagII}
It is worth comparing the new version of \STag with the first version, especially as it has undergone some significant changes. Whilst \STagII can no longer classify Type IIb SN, the remaining 4 types are considered to be stronger classifications due to the changes made to the tags and the inclusion of equivalent widths. The addition of the \textit{r}lap score means that we now have a way of better quantifying whether a classification is trustworthy, with the added bonus of enabling us to give a prediction of the phase. The additional redshift checking is also an improvement for the reliability of \STag. Finally, the fact that we have now trained with not only significantly more spectra than before, but these spectra now also include galaxy light and so are more representative of real data, meaning \STagII does not require near-perfect spectra.

\section{Conclusion}\label{sec:conc}
\STagII is an updated version of the spectroscopic SN classifier \STag, featuring improved tags, more robust classes, and additional functionalities such as phase estimation, equivalent width calculations, and a new measure of trustworthiness from \textit{r}lap scores. Many of these improvements were made possible by moving away from template spectra and using DESI simulated spectra which include both real galaxy and model SN light. \STagII continues to accurately classify a range of spectra, though we emphasise its 99\% accuracy for classifying Type Ia SN, which are of particular importance for cosmology. Furthermore, whilst a test of \STagII on real SNe potentially observed by DESI resulted in accurate classifications in 2 out of 8 cases, we note that this is an issue of completeness and not with accuracy as the remaining 6 that were not classified, could not be due to the \textit{r}lap cut. This criteria was likely failed due to either incorrect redshifts, unique spectra, or being observed well before maximum light. In all cases where \STagII returned a classification, it did so with 100\% accuracy. We also note that cross-checking with TNS discoveries is not how \STag is intended to be used, and instead will be used directly on transients identified by a vision transformer as part of the DESI pipeline \citep{BenZvi2024}.

We are now capable of reporting even more extra information about a given spectrum beyond just tag probabilities, as equivalent widths, phase estimates, and an indication of the accuracy of the associated redshift are now also provided. As more and more data comes in from DESI, we expect \STag will play a vital role in being able to classify any serendipitous SN that are identified. Future work would likely involve expanding and refining the phase bins and spectra used for each bin to improve the \textit{r}lap cut procedure, as this is reliant on having good and complete examples of each type. It would also be prudent to consider expanding to include sub-types that have well-defined unique features and possibly feeding phase information back into the classifier to appropriately deal with lines that are time sensitive.

Finally, we are planning to use \STag as part of a broader part of the DESI operations whereby potential SN spectra observed by DESI will be filtered out and then classified by \STag. These classifications, as well as tag probability values, will then be reported in a possible upcoming value added catalogue that will be part of the larger DESI collaboration as a whole.

\section{Code Accessibility}\label{sec:code}
The code for \STag can be found in the GitHub repository at the following link: \url{https://github.com/wdavison909/STag} which contains all files needed to run \STag, as well as a notebook to demonstrate its use. The spectra used to train the neural network are available upon request.

\clearpage
\appendix
\section{TNS Classification Results}\label{app:tns}

\begin{longtable}{ccccccccccccc}
\caption{Comparison between TNS and \STag classifications for the 8 SN found to be matching between DESI and TNS used in this paper. $\Delta t_{(\mathrm{DESI} - \mathrm{TNS})}$ is the time difference between the date the DESI observation was made and when the TNS spectra \citep{2021TNSCR3766....1S,2021TNSCR4020....1S,2021TNSCR1646....1V,2021TNSCR2204....1S,2021TNSCR2421....1S,2021TNSCR2581....1S,2021TNSCR2659....1P,2021TNSCR3273....1B} was observed. The DESI redshift is as calculated by \texttt{Redrock} whilst the \STag redshift is changed should a more accurate redshift be found following the steps outlined in Section \ref{sec:zcheck}. All tags, regardless of probability, are reported. Only SN 2021aexj and SN 2021qtc passed the \textit{r}lap cut imposed by \STag. \label{tab:tns_class}}
\\
\hline
\hline
 &  & Redshift & Class & \multicolumn{9}{c}{Feature Tags}
\\
& $\Delta t_{(\mathrm{DESI} - \mathrm{TNS})}$ & DESI & TNS & H$\alpha$ & n-H$\alpha$ & H$\beta$ & \CaII & \SiII & \SiII & \SII & \HeI & \FeII
\\
Transient & (Days) & ({\tt STag}) & ({\tt STag}) & & & & H\&K & $\lambda4000$ & $\lambda6355$ & & $\lambda5876$ & $\lambda5170$
\\
\hline
SN 2021acgl & -20 & 0.091 & Ia & 0.10 & 0.91 & 0.02 & 0.59 & 0.71 & 0.39 & 0.83 & 0.71 & 0.81\\
& & (--) & (--) & & & & & & & & &
\\ 
SN 2021aexj & +7 & 0.020 & Ia & 0.75 & 0.38 & 0.00 & 1.00 & 0.79 & 1.00 & 0.02 & 0.99 & 0.04\\
& & (0.051) & (Ia) & & & & & & & & & 
\\
SN 2021ihf & -10 & 0.14 & Ia-pec & 0.66 & 0.07 & 0.01 & 0.07 & 0.12 & 0.50 & 0.77 & 0.98 & 0.32\\
& & (--) & (--) & & & & & & & & & 
\\
SN 2021qtc & +3 & 0.073 & Ia & 0.66 & 0.30 & 0.13 & 1.00 & 0.96 & 1.00 & 1.00 & 0.75 & 0.98\\
& & (0.073) & (Ia) & & & & & & & & &
\\
SN 2021ses & -13 & 0.081 & Ia & 0.85 & 0.11 & 0.87 & 0.00 & 0.18 & 0.01 & 0.45 & 0.95 & 0.60\\ 
& & (--) & (--) & & & & & & & & &
\\ 
SN 2021sxf & -28 & 0.081 & Ia & 0.59 & 0.05 & 0.38 & 0.32 & 0.11 & 0.22 & 0.04 & 0.97 & 0.69\\
& & (--) & (--) & & & & & & & & &  
\\
SN 2021tdl & -28 & 0.046 & II & 0.88 & 0.04 & 0.05 & 0.13 & 0.81 & 0.00 & 0.97 & 0.75 & 0.80\\
& & (--) & (--) & & & & & & & & &   
\\
SN 2021zfs & -1 & 0.68 & Ia & 0.46 & 0.31 & 0.92 & 0.97 & 0.50 & 0.22 & 0.83 & 0.68 & 0.34\\
& & (--) & (--) & & & & & & & & &
\\ 
\hline
\end{longtable}

\section{Classification of SN 2023ixf}\label{app:ixf}
On 19 May 2023 SN 2023ixf was discovered in its host galaxy M101 \citep{2023TNSTR1158....1I}. It was also classified on the same day as a Type II \citep{2023TNSCR1164....1P}, which was matched by subsequent classifications; though some classified it as the sub-type IIn \citep{2023TNSCR1233....1T,2023TNSCR1231....1B,2023TNSCR1267....1Y,2023TNSCR1768....1V}. Due to its extremely close proximity ($z = 0.0008$), SN 2023ixf was also very bright, making it an ideal candidate for observation, which was done over the course of 22 nights from 2023-05-22 to 2023-06-20 by DESI. As such, we were able to make use of the spectra obtained during the course of this observation and use \STag to classify it; the results of these classifications can be found in Table \ref{tab:ixf_class}.

\begin{figure}[htbp]
\centering
\includegraphics[]{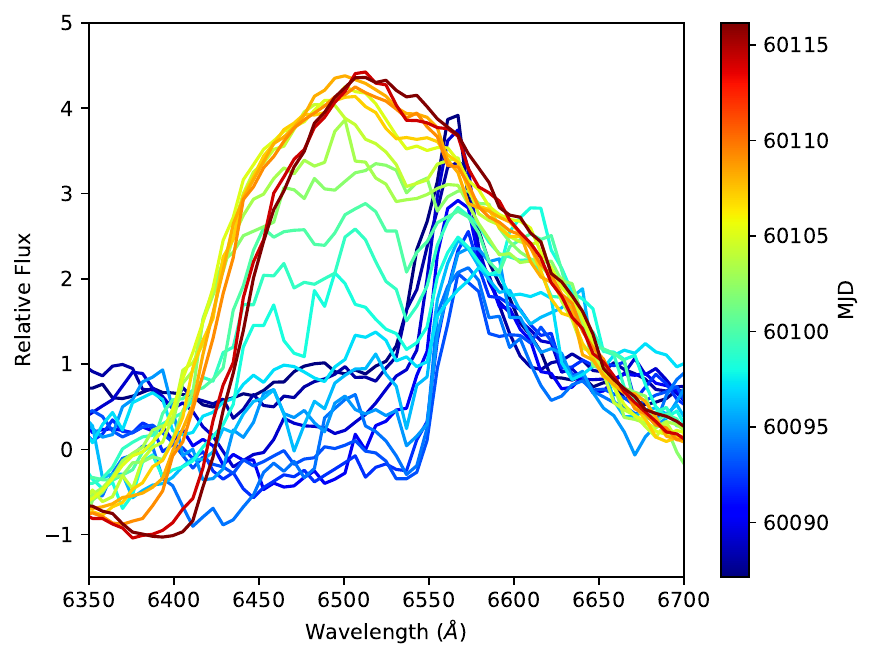}
\caption{Change in the shape of the H$\alpha$ emission line feature of SN~2023ixf over time, with earlier dates in blue changing to later dates in red. The H$\alpha$ line starts out narrow, approximately centered in the associated tag wavelength range (6507 - 6622\AA), but becomes significantly broader to the point where only the right half of the emission line can be seen in that same wavelength range. \label{fig:ha}}
\end{figure}

Most interesting to note is how the two H$\alpha$ tags evolve over the course of observations. During the first week and a half of observations, none of the spectra pass the \textit{r}lap cut and so no classification is made despite the presence of n-H$\alpha$. The implication is that we likely lack suitable Type II templates for these early phases in order to get a good match when cross-correlating, as based purely on the tags alone a Type II classification would be expected. This lack of suitable templates can be explained by the fact that the narrow hydrogen emission is caused by flash ionisation of hydrogen from interactions with some circumstellar medium \citep{Zimmerman2024}, which results in a Type IIn SN \citep{Khazov2016}, of which we do not have any templates included in our list for cross-correlating. The hydrogen feature eventually switches from narrow to a broader feature and we get the expected classification from \STagII (see Figure \ref{fig:ha}).

\clearpage

\begin{longtable}{cccccccccccc}
\caption{Classification results from \STag for spectra observed by DESI of SN 2023ixf. All tags, regardless of probability, are reported. Maximum light is taken to be 2023-05-24 as calculated in \cite{Jacobson-Galan2023}. \label{tab:ixf_class}}
\\
\hline
\hline
& & & \multicolumn{9}{c}{Feature Tags} 
\\
Mean & Days Since & \STag & H$\alpha$ & n-H$\alpha$ & H$\beta$ & \CaII & \SiII & \SiII & \SII & \HeI & \FeII
\\
Julian Date & Maximum Light & Class & & & & H\&K & $\lambda4000$ & $\lambda6355$ & & $\lambda5876$ & $\lambda5170$
\\
\hline
60086 & -2 & -- & 0.52 & 0.99 & 1.00 & 0.13 & 0.64 & 0.24 & 0.70 & 0.98 & 0.44\\
60087 & -1 & -- & 0.50 & 0.99 & 0.98 & 0.22 & 0.92 & 0.06 & 0.18 & 0.94 & 0.47\\
60088 & 0 & -- & 0.12 & 1.00 & 0.34 & 0.38 & 0.61 & 0.39 & 0.16 & 0.87 & 0.33\\
60089 & +1 & -- & 0.06 & 1.00 & 0.30 & 0.13 & 0.42 & 0.16 & 0.01 & 0.63 & 0.20\\
60091 & +3 & -- & 0.09 & 0.99 & 0.13 & 0.52 & 0.94 & 0.45 & 0.00 & 0.67 & 0.55\\
60092 & +4 & -- & 0.20 & 0.95 & 0.75 & 0.26 & 0.28 & 0.12 & 0.00 & 0.84 & 0.21\\
60093 & +5 & -- & 0.17 & 0.98 & 0.08 & 0.87 & 0.68 & 0.10 & 0.02 & 0.31 & 0.12\\
60094 & +6 & -- & 0.41 & 0.92 & 0.01 & 0.06 & 0.63 & 0.03 & 0.01 & 0.36 & 0.34\\
60095 & +7 & -- & 0.82 & 0.62 & 0.22 & 0.08 & 0.93 & 0.08 & 0.01 & 0.57 & 0.11\\
60096 & +8 & -- & 0.96 & 0.06 & 0.54 & 0.06 & 0.98 & 0.09 & 0.03 & 0.96 & 0.34\\
60097 & +9 & -- & 1.00 & 0.02 & 0.04 & 0.03 & 1.00 & 0.10 & 0.01 & 0.38 & 0.23\\
60098 & +10 & -- & 1.00 & 0.03 & 0.05 & 0.02 & 0.98 & 0.02 & 0.01 & 0.60 & 0.31\\
60099 & +11 & -- & 1.00 & 0.00 & 0.01 & 0.02 & 0.99 & 0.01 & 0.05 & 0.30 & 0.13\\
60101 & +13 & II & 1.00 & 0.00 & 0.03 & 0.00 & 0.99 & 0.03 & 0.43 & 0.02 & 0.22\\
60102 & +14 & II & 1.00 & 0.00 & 0.06 & 0.00 & 0.96 & 0.03 & 0.64 & 0.01 & 0.28\\
60103 & +15 & II & 1.00 & 0.00 & 0.04 & 0.00 & 0.94 & 0.09 & 0.67 & 0.01 & 0.26\\
60104 & +16 & II & 1.00 & 0.00 & 0.04 & 0.00 & 0.95 & 0.07 & 0.37 & 0.03 & 0.16\\
60106 & +18 & II & 1.00 & 0.00 & 0.03 & 0.00 & 0.91 & 0.03 & 0.42 & 0.01 & 0.22\\
60107 & +19 & II & 1.00 & 0.00 & 0.04 & 0.00 & 0.95 & 0.08 & 0.27 & 0.01 & 0.23\\
60108 & +20 & II & 1.00 & 0.00 & 0.05 & 0.00 & 0.91 & 0.07 & 0.38 & 0.01 & 0.19\\
60113 & +25 & II & 1.00 & 0.00 & 0.05 & 0.00 & 0.90 & 0.14 & 0.17 & 0.03 & 0.24\\
60115 & +27 & II & 1.00 & 0.00 & 0.05 & 0.00 & 0.88 & 0.14 & 0.16 & 0.04 & 0.18\\
\hline
\end{longtable}


\begin{acknowledgments}

This approach was inspired by a presentation given by Lawrence Rudnick. WD and DP are supported by the project ``Understanding Dark Universe Using Large Scale Structure of the Universe'', funded by the Korean Ministry of Science. 

This material is based upon work supported by the U.S. Department of Energy (DOE), Office of Science, Office of High-Energy Physics, under Contract No. DE–AC02–05CH11231, and by the National Energy Research Scientific Computing Center, a DOE Office of Science User Facility under the same contract. Additional support for DESI was provided by the U.S. National Science Foundation (NSF), Division of Astronomical Sciences under Contract No. AST-0950945 to the NSF’s National Optical-Infrared Astronomy Research Laboratory; the Science and Technology Facilities Council of the United Kingdom; the Gordon and Betty Moore Foundation; the Heising-Simons Foundation; the French Alternative Energies and Atomic Energy Commission (CEA); the National Council of Humanities, Science and Technology of Mexico (CONAHCYT); the Ministry of Science and Innovation of Spain (MICINN), and by the DESI Member Institutions: \url{https://www.desi.lbl.gov/collaborating-institutions}. Any opinions, findings, and conclusions or recommendations expressed in this material are those of the author(s) and do not necessarily reflect the views of the U. S. National Science Foundation, the U. S. Department of Energy, or any of the listed funding agencies.

The authors are honored to be permitted to conduct scientific research on Iolkam Du’ag (Kitt Peak), a mountain with particular significance to the Tohono O’odham Nation.

\end{acknowledgments}

\software{\texttt{Astropy} \citep{astropy:2013,astropy:2018},
\texttt{DASH} \citep{Muthukrishna2019},
\texttt{keras} \citep{chollet2015keras},
\texttt{Matplotlib} \citep{Hunter:2007},
\texttt{NumPy} \citep{harris2020array},
\texttt{scikit-learn} \citep{scikit-learn},
\texttt{SciPy} \citep{2020SciPy-NMeth},
\texttt{STag} \citep{Davison2022}
\texttt{TensorFlow} \citep{tensorflow2015-whitepaper}
}


\bibliography{references}{}
\bibliographystyle{aasjournalv7}

\end{document}